\documentclass[letter]{aa} 

\usepackage{graphicx}
\usepackage[]{natbib}
\usepackage{txfonts}
\usepackage[utf8]{inputenc}
\newcommand{\teff}{T$_{\rm eff}$}
\newcommand{\logg}{$\log~g$}
\newcommand{\feh}{$\rm [Fe/H]$}
\newcommand{\met}{${\rm [M/H]}$}
\newcommand{\aabun}{${\rm [\alpha/Fe]}$}
\newcommand{\mgfeabun}{${\rm [Mg/Fe]}$}
\newcommand{\sifeabun}{${\rm [Si/Fe]}$}
\newcommand{\cafeabun}{${\rm [Ca/Fe]}$}
\newcommand{\alfeabun}{${\rm [Al/Fe]}$}

\newcommand{\mgsicafeabun}{${\rm [Mg+Si+Ca/Fe]}$}

\begin{document} 

 \title{The Gaia-ESO Survey: low-$\alpha$ element stars in the Galactic Bulge\thanks{Based on data products from observations made with ESO Telescopes at the La Silla Paranal Observatory under programme ID 188.B-3002. These data products have been processed by the Cambridge Astronomy Survey Unit (CASU) at the Institute of Astronomy, University of Cambridge, and by the FLAMES/UVES reduction team at INAF/Osservatorio Astrofisico di Arcetri. These data have been obtained from the Gaia-ESO Survey Data Archive, prepared and hosted by the Wide Field Astronomy Unit, Institute for Astronomy, University of Edinburgh, which is funded by the UK Science and Technology Facilities Council.}}
\author{A. Recio-Blanco\inst{1}
          \and
          A. Rojas-Arriagada\inst{1}
   	 \and
          P. de Laverny\inst{1}
          \and
          S. Mikolaitis\inst{2}
          \and
          V. Hill\inst{1}
          \and
          M. Zoccali\inst{3,4}
          \and
          J.G. Fern\'andez-Trincado\inst{5}
          \and
          A. C. Robin\inst{5}
          \and
          C. Babusiaux\inst{6}
          \and
          G. Gilmore\inst{7}
          \and
          S. Randich\inst{8}
          \and
          E. Alfaro\inst{9}
          \and
          C. Allende Prieto\inst{10}
          \and
          A. Bragaglia\inst{11}
          \and
          G. Carraro\inst{12}
          \and
          P. Jofr\'e \inst{7}\inst{13}
          \and
          C. Lardo\inst{14}
          \and
          L. Monaco\inst{15}
          \and
          L. Morbidelli\inst{8}
          \and
          S. Zaggia\inst{16}
          }
\institute{
         Universit\'e C\^ote d'Azur, Observatoire de la C\^ote d'Azur, CNRS, Laboratoire Lagrange, France \\  \email{arecio@oca.eu}
         \and
         Institute of Theoretical Physics and Astronomy, Vilnius University, A. Go$\check{s}$tauto 12, 01108 Vilnius, Lithuania
         \and
         Instituto de Astrof\'isica, Facultad de F\'isica, Pontificia Universidad Cat\'olica de Chile, Av. Vicu\~na Mackenna 4860, Santiago, Chile
         \and 
	Millennium Institute of Astrophysics, Av. Vicu\~na Mackenna 4860, 782-0436 Macul, Santiago, Chile
         \and 
         Institut Utinam, CNRS, Universit\'e de Franche-Comt\'e, OSU THETA Franche-Comt\'e-Bourgogne, Obs. de Besan\c{c}on, France.
         \and
         GEPI, Observatoire de Paris, CNRS, Universit\'e Paris Diderot, 5 Place Jules Janssen, 92190 Meudon, France
          \and
         Institute of Astronomy, Cambridge University, Madingley Road, Cambridge CB3 0HA, United Kingdom 
         \and
         INAF-Osservatorio Astrofisico di Arcetri, Largo E. Fermi, 5, 50125, Firenze, Italy
         \and
         Instituto de Astrof\'{i}sica de Andaluc\'{i}a-CSIC, Apdo. 3004, 18080, Granada, Spain
         \and
         Instituto de Astrof\'{\i}sica de Canarias, E-38205 La Laguna, Tenerife, Spain
         \and
         INAF - Osservatorio Astronomico di Bologna, via Ranzani 1, 40127, Bologna, Italy
         \and
         Dip. di Fisica e Astronomia, Univ. de Padova, Vicolo Osservatorio 3, 35122 Padova, Italy
         \and
         N\'ucleo de Astronom\'ia, Facultad de Ingenier\'ia, Universidad Diego Portales,  Av. Ejercito 441, Santiago, Chile
         \and
         Astrophysics Research Institute, Liverpool John Moores University, 146 Brownlow Hill, Liverpool L3 5RF, United Kingdom
         \and
         Departamento de Ciencias Fisicas, Universidad Andres Bello, Republica 220, Santiago, Chile
         \and
         INAF Padova Observatory, Vicolo dell'Osservatorio 5, 35122 Padova, Italy
         }
   \date{Received ...; accepted ...}
   \abstract {We take advantage of the Gaia-ESO Survey iDR4 bulge data to search for abundance anomalies that could
shed light on the composite nature of the Milky Way bulge. The $\alpha$-elements \ (Mg, Si, and whenever available, Ca) abundances, and their trends with Fe abundances have been analysed for a total of 776 bulge stars. In addition, the aluminum abundances and their ratio to Fe and Mg have also been examined. Our analysis reveals the existence of low-$\alpha$ element abundance stars with respect to the standard bulge sequence in the \aabun \ vs. \feh \ plane. 18 objects present deviations in \aabun \  ranging from 2.1 to 5.3 $\sigma$ with respect to the median standard value. Those stars do not show Mg-Al anti-correlation patterns. Incidentally, this sign of the existence of multiple stellar populations is reported firmly for the first time for the bulge globular cluster NGC~6522. The identified low-$\alpha$ abundance stars have chemical patterns compatible with those of the thin disc. Their link with massive dwarf galaxies accretion seems unlikely, as larger deviations in $\alpha$ abundance and Al would be expected. The vision of a bulge composite nature and a complex formation process is reinforced  by our results. The used approach, a multi-method and model-driven analysis of high resolution data seems crucial to reveal this complexity.}

\maketitle
   
%

\section{Introduction}
The study of the Galactic bulge is rapidly unveiling its complexity. 
Because of its physical properties (metallicity distribution, age, spatial location, kinematical
features...), the bulge is at the cross-road of the other main Galactic components like the halo, the thick disc and the thin disc.
As a consequence, the formation scenarios currently invoked are directly linked to the general evolution of the Milky Way
and to one crucial opened question: the importance of fast versus secular evolution.
Three main scenarios are proposed: i) in situ formation via dissipative collapse of a protogalactic gas cloud \citep{ELS62}, 
ii) accretion of substructures in a $\Lambda$CDM context \citep[e.g.][]{ScannapiecoTissera03}, and (iii) secular formation from disc material 
through bar formation, vertical instability, buckling, and fattening, producing a boxy/peanut bulge 
\citep[e.g.][]{MartinezValpuesta13, DiMatteo14}.

Our knowledge of the bulge has improved relevantly in the recent years, however, many key opened questions remain.
By way of example, the number of bulge components, estimated from the metallicity distribution function, kinematical
and structural data,  is still under debate. Although up to five components have been suggested \citep[e.g.][]{NessPop}, 
recent studies \citep[e.g.][Zoccali et al. 2016]{Alvaro1, Mathias16} show that only two components seem to be necessary. 
In this context, precise individual chemical abundances are crucial to disentangle the different evolutionary pathways responsible 
for the bulge formation. The first studies in this sense regarded mainly (but not exclusively) $\alpha$-element abundances 
\citep[e.g.][]{McWilliamRich, Bensby13, Oscar15}
that, combined with iron abundances, can unveil important information regarding the initial mass function and the star formation history of
a stellar system. Up to now, the different analyses have revealed a single sequence in the \aabun \ vs. \feh \ plane, flattening at metallicities lower than \mbox{-0.37$\pm$0.09 dex}
\citep[e.g.][]{Alvaro2}, the metallicity at which the maximum of supernovae type Ia rate occurred. However, recent observations extending the analysis to other elements have already 
detected departures from what seemed to be a simple chemical evolutionary path, like the existence of nitrogen over-abundant stars 
\citep[][]{Schiavon16}.

The goal of this paper is to take advantage of the Gaia-ESO Survey \citep[GES,][]{GES, GESRandich} bulge data to search for abundance anomalies that could
reveal the bulge composite nature. The data are described in Sect. 2, our results are presented in Sect. 3 and their interpretation
is developed in the final Sect. 4.

\section{GES stellar parameters, distances and orbits.}

We have used the atmospheric parameters and individual chemical abundances (Fe, Mg, Si and Al) of bulge stars observed by GES, using the GIRAFFE 
spectrograph, and included in its fourth internal data release. 
The data are described in detail in \cite{Alvaro2}. The initial total sample comprises 2320 red clump stars in 11 bulge fields, sampling the area
-10$^{\circ} \le$ {\it l} $\le$ +10$^{\circ}$ and 10$^{\circ} \le$ {\it b} $\le$ -4$^{\circ}$.

We remind that the HR21 setup was employed for all the bulge stars.
In addition, 172 stars in Baade's Window were observed with the HR10 setup. Those stars are in common with the analysis of \cite{Van11}.

As the search for abundance anomalies has to rely on very precise results, we have applied 
the following selection criteria to the initial sample:  2.0$<$ \logg$<$3.0~dex, $\sigma_{Teff}<$250~K, $\sigma_{\log \rm g}<$0.6~dex, $\sigma_{\rm [Fe/H]}<$0.1~dex, $\sigma_{\rm [Mg/H]}<$0.1 dex, $\sigma_{\rm [Si/H]}<$0.1 dex, $\sigma_{\rm [Al/H]}<$0.1 dex. This reduces the sample to 776 stars
with very high quality parameters (42 of them observed with two setups). The mean signal-to-noise of the working sample is 280$\pm$70 (deviation estimated from the MAD).
We make use of the \cite{Alvaro2} spectroscopic distances, estimated for the same data set, to ensure that our selection is confined 
to the bulge region (R$_{GC}<3.5$ kpc). The median error of the distances is 1.2 kpc (around 16\%). Finally, OGLE proper motions available for the subsample of stars in Baade's Window \citep[][]{Sumi} allowed the
estimation of the stellar orbital parameters. To this purpose, we have adopted the Model 4 presented in \cite{JoseModel} and  composed of nonaxisymmetric potentials.

\section{Search for $\alpha$-element abundance anomalies}
The main challenge of testing the existence of stars with non standard low-$\alpha$ element abundances is demonstrating 
that they do not belong to the high error queue of a normal abundance distribution. 
To this purpose, we have first 
d a fiducial median profile and a 1$\sigma$ dispersion
band, over 11 iron abundance bins for the \mgfeabun, the \sifeabun \ and the \aabun \ abundances. 
Secondly, we have computed, for each star, the difference in  \mgfeabun, \sifeabun \ and \aabun \ with respect to
the median values of the corresponding iron bin (in the sense median minus sample),
called $\Delta$\mgfeabun, $\Delta$\sifeabun \ and $\Delta$\aabun, respectively.

If there are low-$\alpha$ element stars in the bulge, the distribution of the $\Delta$\aabun \ around the standard value should
be 1) skewed, and 2) not corresponding to a single gaussian.  We have first applied a D'Agostino skewness test
to  the $\Delta$\aabun \ abundance distribution. 
The skewness value of the $\Delta$\aabun \ is 47.5 (p-value of 4.88e-11), confirming the asymmetry. 
We have tested the robustness of this result to the existence of strong outliers by reapplying the D'Agostino test considering
only the stars within 2 sigma from the median  \aabun \ abundance value. Both the z-score of the test (6.9) and its p-value
(0.03) seem to confirm that even the core of the distribution deviates from normality and it is skewed towards the low-alpha side.
In addition, 
we have applied a Gaussian Mixture Method \citep[GMM,][]{AlvaroDisc} to estimate the number of components of the distribution. The 
GMM algorithm constructs a generative model that consists in the specific Gaussian 
mixture that better predicts the data structure. We adopted the Akaike information criterion (AIC) as a cost function to assess the relative fitting quality between 
different proposed mixtures. 
The results of this analysis are shown in Figure 1. Clearly, the single component solution can be excluded to explain the data distribution, highlighting the existence of an asymmetry.

Furthermore, the dispersion of the measured \aabun \ abundances in metal-rich globular cluster giants can help us to realistically evaluate the errors on the bulge abundances.
 We have considered two observed clusters, within the same constraints as in our bulge analysis sample (c.f. Sect. 2): NGC~104  (\feh=-0.66$\pm$0.05~dex, 30 stars) and NGC~5927 (\feh=-0.33$\pm$0.06~dex, 56 stars). The two clusters dispersion in  the \aabun \ abundances (0.035~dex  and 0.045~dex, respectively), the \mgfeabun (0.04 and 0.05~dex, respectively) and the \sifeabun (0.05 and 0.06~dex, respectively) 
is clearly lower than that of the bulge distribution (0.06~dex for \aabun, 0.07~dex for \mgfeabun and 0.08~dex for \sifeabun), reinforcing the hypothesis of a complex \aabun \ v.s. \feh \ sequence in the bulge.

\begin{figure}[ht]
\includegraphics[width=9cm,height=5cm]{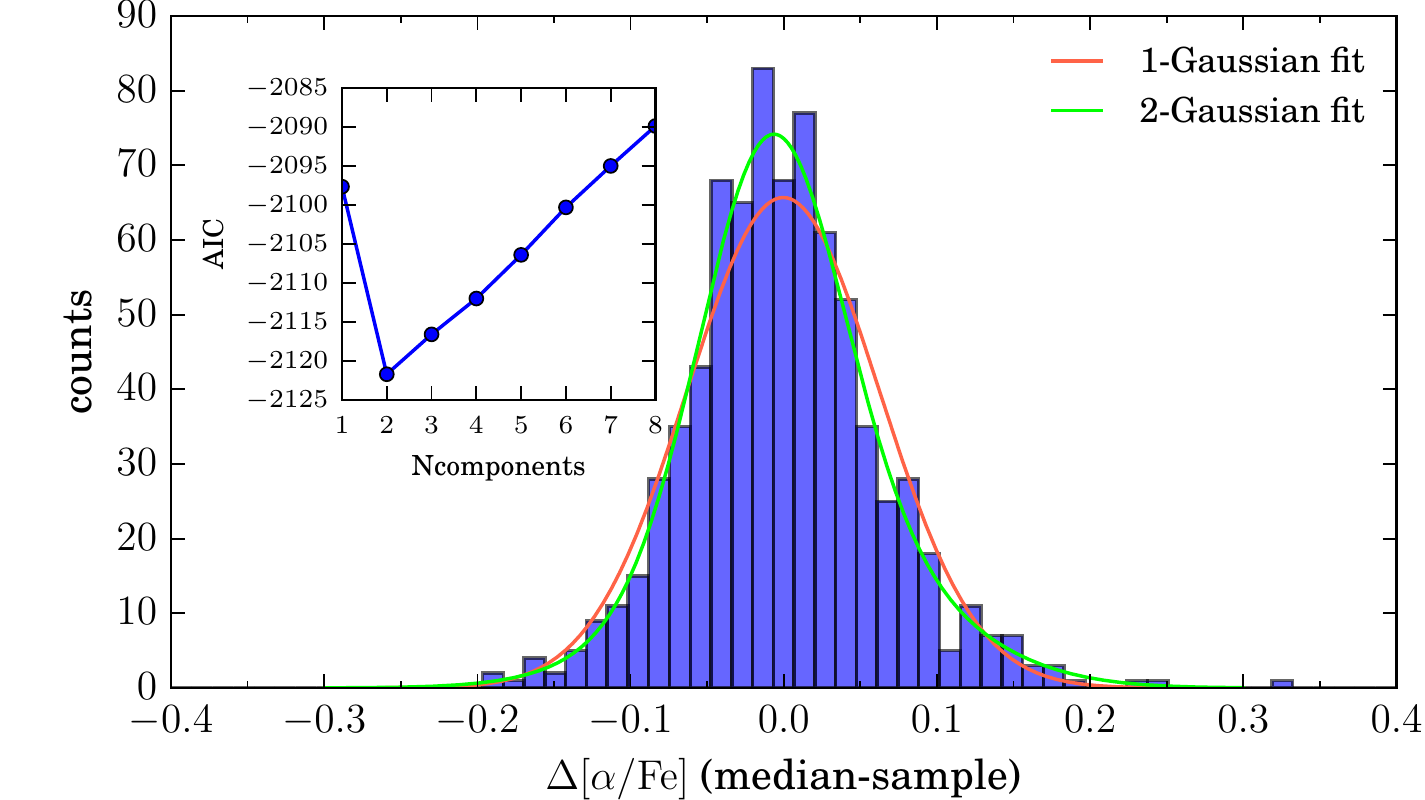}
\caption{GMM analysis of the  \aabun \ abundance distribution with respect to the median values along the sequence. The inner panel shows the Akaike information criterion for different numbers of components. The best fit (green curve) corresponds to the two-component mixture. The red curve shows the poorer quality fit of the single gaussian component (its probability of explaining the distribution is 21 times lower than the double gaussian one).}
\label{BW}
\end{figure}

\subsection{Stars selection separately from different $\alpha$-elements.}

First of all, we have analyzed the abundances of Mg and Si, and their trends with the Fe abundance.
Figure 2 shows the \mgfeabun \ vs. \feh \ distribution (upper panel), the \sifeabun \ vs. \feh \ one (middle panel) and the \aabun \ vs. \feh \ distribution defined as 
(\mgfeabun+\sifeabun)/2 (lower panel). Then, we have identified the stars showing deviations larger than 1.2 $\sigma$ in $\Delta$\mgfeabun \ and in $\Delta$\sifeabun \ (colour coded in the upper and middle panels
of Fig. 2). To ensure that the \aabun \ anomaly is present in more than one element \citep[avoiding possible Mg depleted globular cluster \textit{escap\'ees}, e.g.][]{Eugenio09}, we have restricted the selection to the intersection of the low-Mg and low-Si subsamples.
Finally, an additional condition was imposed: only the objects showing deviations greater than 2 $\sigma$ in $\Delta$\aabun \ were selected.
The final subsample of stars are colour coded in the bottom panel of Fig.~2. It includes 18 objects 
with \aabun \ underabundances ranging from 2.1 to 5.3 $\sigma$. 90\% of them are within 2.5~kpc from the Galactic centre and two thirds are within 1.8~kpc.

\begin{figure}[h]
\includegraphics[width=10cm,height=9cm]{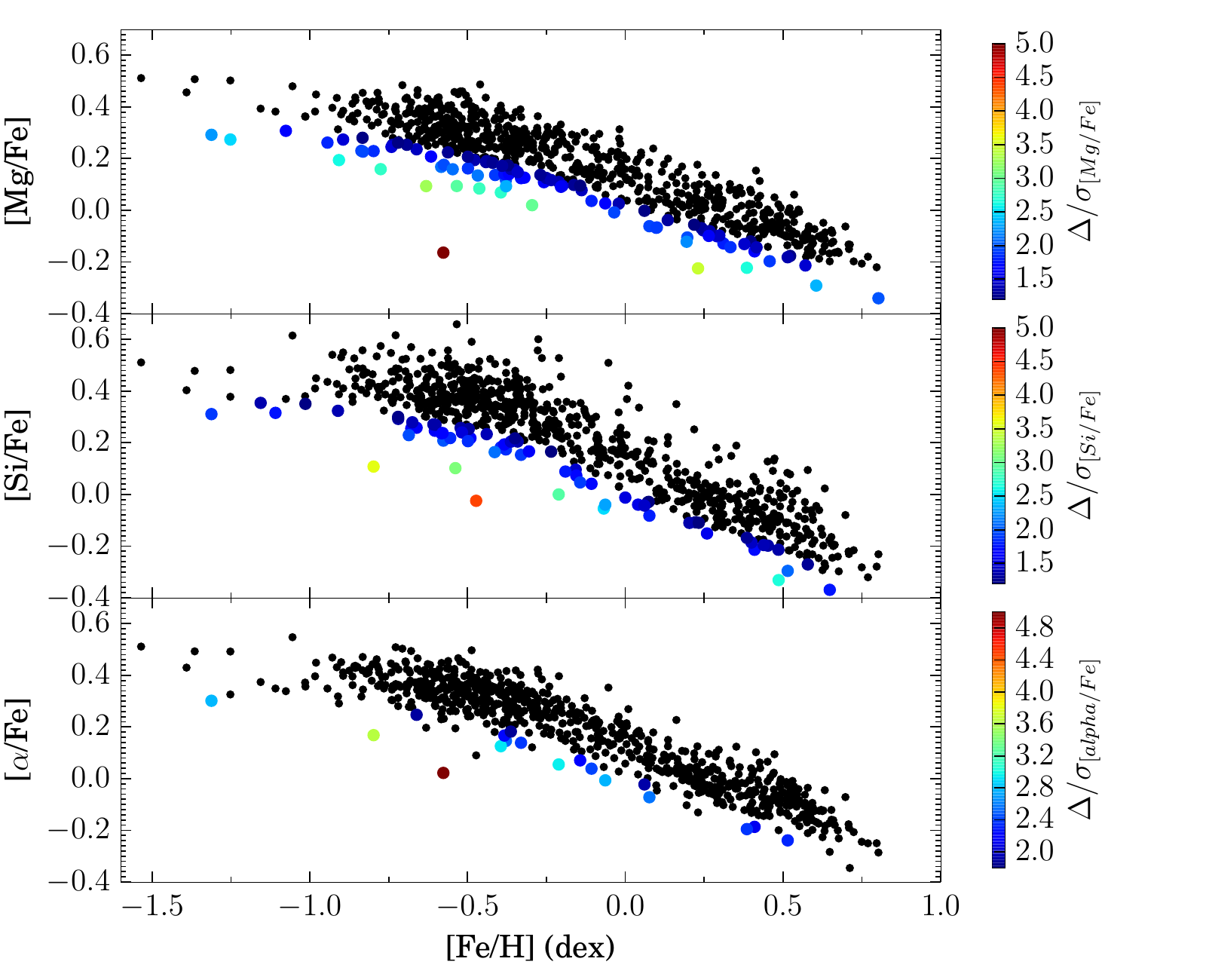} 
\caption{Bulge distribution of \mgfeabun, \sifeabun \ and \aabun  \ with respect to \feh \ (upper, middle and lower panels, respectively). 
The colour code shows the difference 
with respect to the median value ($\Delta$\mgfeabun, $\Delta$\sifeabun \ and $\Delta$\aabun ) expressed in units of standard deviation.
Both $\Delta$ and $\sigma$ are evaluated as a function of \feh.}
\label{MgFe}
\end{figure}

A number of checks have been performed to further test the reliability of the abundance anomalies. First of all, it has been verified that the measurement dispersions  in the parameters 
and the abundances provided by the different GES analysis nodes (and taken as a proxy of the error) are similar for the anomalous stars and for the reference sample of 776 objects. 
In addition, the spectra of representative stars showing 
anomalies have been compared to those of stellar twins (with maximum differences of 50 K for \teff,
0.02 dex for \logg \ and 0.03 dex for \feh). This is possible thanks to the fact that the bulge GES targets are restricted to the red clump and the chances of finding
stellar twins in the sample are high. These comparisons allowed
to confirm that the Mg and Si abundances of the identified anomalous stars are in fact underabundant with respect to their twins of standard composition. As an example, the spectrum of
the anomalous star 17553025-4106299 (with \mgfeabun$=$-0.16$\pm$0.08 dex, \sifeabun$=$0.21$\pm$0.08 dex) differs clearly in the $\alpha$-element lines from that of its twin 18262565-3151577, having \mgfeabun$=$0.31$\pm$0.09 dex and \sifeabun$=$0.37$\pm$0.06 dex. An additional illustration of spectral twins comparison is given in Figure B.1 of the Appendix.
Third, the differences between the spectroscopic and the photometric \teff \ values \citep[derived following][]{GHB09}
of the stars presenting  \aabun \ abundance anomalies have been checked and compared to those of the stars with standard compositions. No particular problems
that could indicate specific errors in the spectroscopic \teff \ of the anomalous stars  have been found. Moreover, the sensitivity to possible \logg \ errors of the Mg and Si lines has
been verified using synthetic spectra. 
Although the estimated Si abundances are slightly sensitive to \logg \ uncertainties,
it is not the case of the Mg abundances. In the same line, the variation of the Mg and Si abundances with the typical  \met \ uncertainties
are at least 6 times smaller than the detected anomalies. Finally, it has been checked that the stars presenting abundance anomalies do not belong to a unique GES
bulge field and GIRAFFE exposure, which could reveal possible problems with the data reduction, like sky sustraction residuals. 
In conclusion, all the above mentioned verification tests confirm the real $\alpha$-poor nature of the identified bulge stars.

\subsection{Baade's Window data}

\begin{figure}[ht]
\includegraphics[width=9cm,height=5cm]{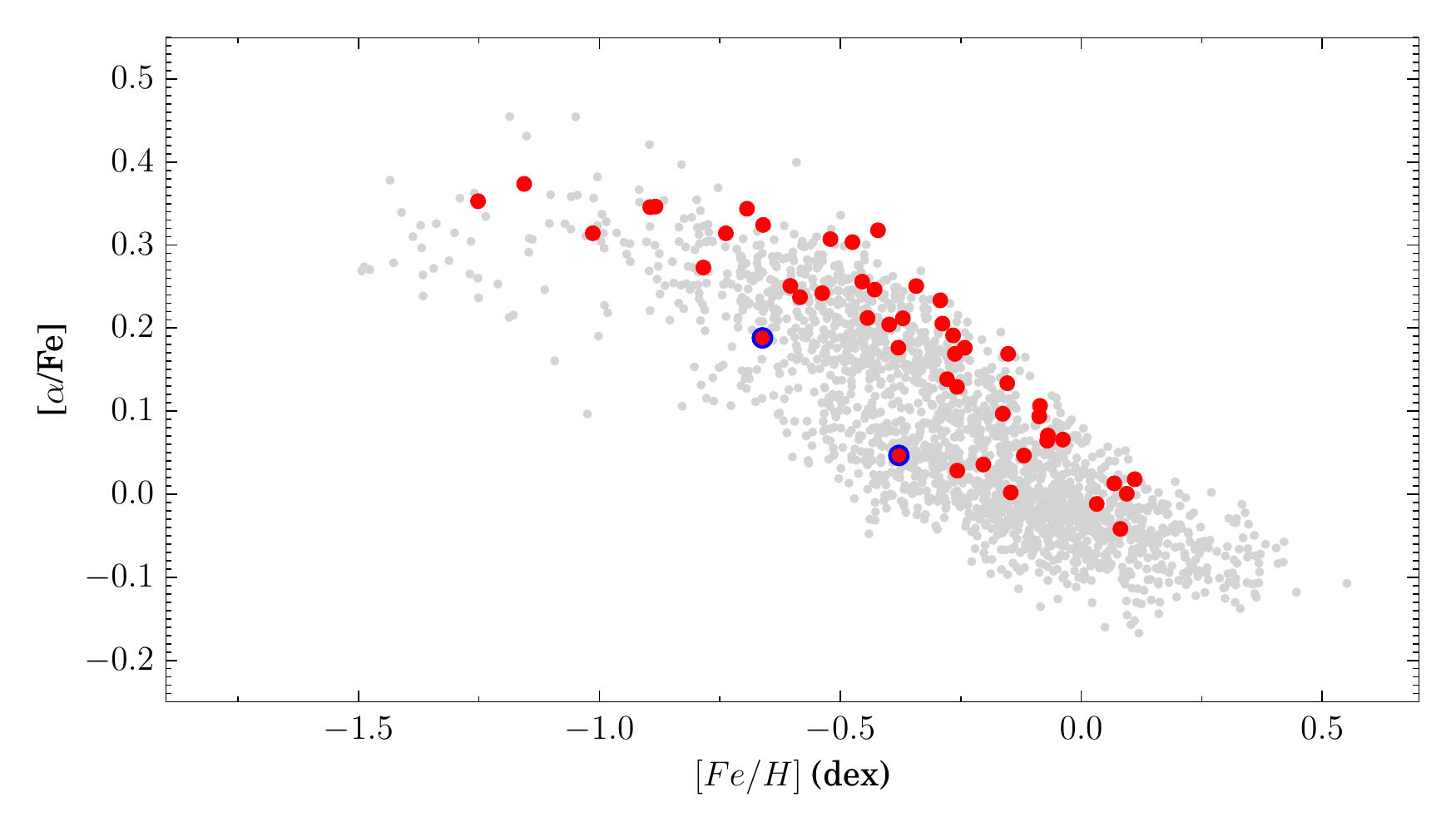} 
\caption{Distribution of \aabun \  vs. \feh \  
for the 48 stars of Baade's Window with high quality estimations of Mg, Si and Ca abundances (red points). The two stars highlighted with blue circles
are part of the group of stars identified to have Mg and Si underabundances in Fig. 2. Grey points
show the \aabun \ vs. \feh \ distribution of GES disc stars.}
\label{BW}
\end{figure}

Particular attention has been paid to the stars in Baade's Window (BW), for which both GES HR21 and HR10 spectra are available.
Those stars, with more reliable parameters and abundance estimations, have measurements of an additional $\alpha$ element (calcium), and they have 
literature \mgfeabun \  abundance determinations from \cite{Van11} for comparison. Figure 3 shows the distribution of  \aabun \  (defined as \mgsicafeabun) vs. \feh \ 
for the 48 stars of BW with high quality estimations of Mg, Si and Ca abundances (red points). The two red circles highlighted with blue edges 
correspond to the stars 18034317-3006349 (\feh$=-0.38\pm0.03$ dex) and 18032412-3003001 (\feh$=-0.66\pm0.02$ dex), already identified to have Mg and Si underabundances
in Fig. 1. 
These two stars are confirmed to be  \aabun \ under abundant, even when Ca is considered. In addition, the star 18034317-3006349 
seems to lay on the locus of the corresponding thin disc sequence, well below the standard \aabun \ values of the bulge at
its metallicity interval. 

Literature values from \cite{Van11} are available for these two stars. First, the 
GES iron abundances are compatible within the errors with the \cite{Van11} ones. Second, the \mgfeabun \  
ratio of \cite{Van11} for the star 18032412-3003001  (0.09 $\pm$ 0.18 dex)  is also 1 $\sigma$ below
the corresponding median \mgfeabun \  value for that metallicity interval. 
They do not report Mg abundance estimations for the other star. 

Lastly, the orbits of the two stars seem confined to the bulge region.
The derived radial apocentric distances for 18034317-3006349 and 18032412-3003001 are 2.75$\pm$1.1~kpc and 1.87$\pm$1.3~kpc respectively.
The corresponding maximum vertical amplitudes are 1.16$\pm$0.2~kpc and 1.66$\pm$0.4~kpc.

\begin{figure}[ht]
\includegraphics[width=9.5cm,height=5cm]{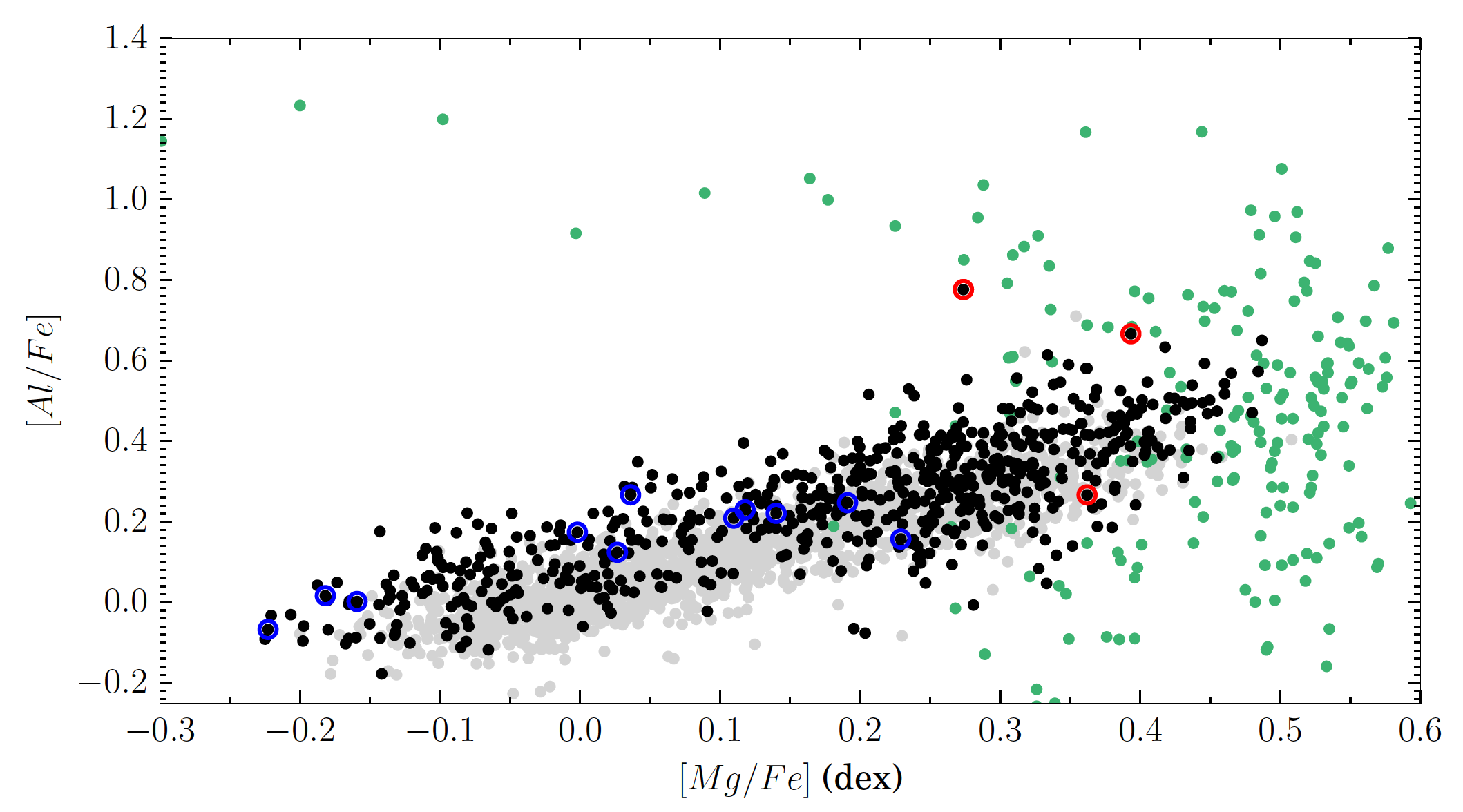} 
\caption{\alfeabun \ abundances with respect to the \mgfeabun \ ratio for bulge stars (black points). Low-$\alpha$ stars and NGC6522 stars
are highlighted with blue and red circles, respectively. Disc stars and globular cluster values from \cite{Eugenio09} are shown as
grey and green points.}
\label{AlFe}
\end{figure}

\subsection{Aluminum abundances}

Two aluminum lines in the HR21 spectra allowed the derivation of Al abundances for the GES bulge data. 
Figure 4 shows the \alfeabun \ abundance with respect to the \mgfeabun \ one for the studied bulge stars (black points). Low-$\alpha$ stars
of Section 3.1 are highlighted with blue circles. As in Fig.~3, the abundances of disc stars are also included for comparison as light grey points.
Finally, green points show the \cite{Eugenio09}  values for GC stars (c.f their Figure~5). 
First, the results presented in Fig. 4~show that the identified low-$\alpha$ bulge stars do not present GC abundance patterns. Their Al abundance seems to be in agreement with the Mg one,
laying on the same sequence as bulge stars of standard composition. Moreover, the disc and the bulge sequences overly in the same locus of the figure, as expected.

Incidentally, three members of the globular cluster NGC~6522 (\feh=-1.14$\pm$0.10~dex) were observed in the GES BW fields (marked with 
red circles in Fig.~4). One of these stars, 18033857-3002434, presents a clear Al  over-abundance with \alfeabun=0.78$\pm$0.03~dex and a
Mg-depletion with respect to the other two (\mgfeabun=0.27$\pm$0.03~dex), with no sign of Si under-abundance (\sifeabun=0.38$\pm$0.07)
or Ca under-abundance (\cafeabun=0.40$\pm$0.07). 
Figure B.2 shows the fit of the two Al lines at 8772.8 and 8773.8 \AA \ for this star, confirming the reliability of the estimated Al abundance and the continuum 
normalisation.  As a consequence, the three NGC~6522 members present clearly a Mg-Al anti-correlation, as seen in Fig.~4. This result 
confirms the previous abundance measurements of  \cite{Ness14}  of 8 potential NGC~6522 members,  and it reinforces the existence of multiple populations
in this old cluster of relatively low-mass.

\section{Discussion}
The analysis of the Gaia-ESO survey bulge data reveals the existence of low-$\alpha$  abundance
stars. 
It is necessary to invoke a complex bulge phylogenesis to explain the reported intrinsic spread of  \aabun  \ abundances at a given metallicity.
The results presented here are compatible with the existence of a (at least) bimodal distribution of \aabun. They also present a more detailed vision of the \aabun \ vs. \feh \
bulge trends than previous works analyzing large samples of stars \citep[e.g.][]{NessPop}, thanks to the GES high resolution and high signal-to-noise
data. They bring into light the presence of bona fide low-$\alpha$ stars in the bulge that were already visible, without being remarked, in several
studies in the literature \citep[e.g.][and references therein]{AnaElia13}.

Several scenarios can be invoked to interpret this situation. First, only massive dwarf galaxies, like the 
Sagittarius dwarf spheroidal, could have benefited from an efficient enough chemical enrichment to reach the metallicity range concerned by our analysis.
In those cases, we would expect deviations in the \aabun \ vs. \feh  \ plane of at least 3$\sigma$ or more
with respect to the standard \aabun \ values of the bulge \citep[e.g.][]{McWilliam13} 
Therefore, very few objects of our study 
have chemical patterns compatible with this explanation.

Second, the identified low-$\alpha$ abundance stars have chemical patterns compatible with those of the thin disc.
A disc contribution to the bulge formation \citep[e.g.][]{DiMatteo15} would indeed leave its imprints in the chemical abundances,
being compatible with the observed outliers from the standard bulge sequence. 
In this sense, a possible spread (or bimodality) of
the stellar ages \citep[][]{Misha16} in the bulge would be in agreement with this scenario. Nevertheless, the  \aabun \ spread
could be present even at metallicities as low as -0.5~dex, challenging this interpretation.

On the other hand, 
our finding of multiple stellar populations in the cluster NGC6522, added to the \cite{Schiavon16b} results for other clusters, reinforces the
link between the globular clusters and the nitrogen abundance anomalies of field stars reported by \cite{Schiavon16}. 

Finally, we would like to point out that the search of non-standard objects benefits from the methodology implemented by GES. First, the use 
of a multi-method spectral analysis allows reducing the inevitable methodological errors. Moreover, the simultaneous 
analysis of several elements having similar nucleosynthetic
origins is important to avoid errors affecting the lines of a particular element. Lastly, model-driven methods like the ones employed in this
study are more appropriate for the search of unexpected objects than data driven methods, working by construction
in a close environment of knowledge.
 
 In summary, the bulge composite nature and, therefore, its complex formation process is reinforced  by the present results, with all the
strength of a multi-method abundance analysis of high resolution and high quality data.

\begin{acknowledgements}
ARB, PdL and VH acknowledge financial support form the ANR 14-CE33-014-01.
This work was partly supported by the European Union FP7 programme through ERC grant number 320360 and by the Leverhulme Trust through grant RPG-2012-541. 
We acknowledge the support from INAF and Ministero dell'Istruzione, dell'Universit\`a e della Ricerca (MIUR) in the form of the grant "Premiale VLT 2012". The results 
presented here benefit from discussions held during the Gaia-ESO workshops and conferences supported by the ESF (European Science Foundation) through the 
GREAT Research Network Programme. M. Zoccali gratefully acknowledge support by the Ministry of Economy, Development, and Tourism's Millenium 
Science Initiative through grant IC120009, awarded to the Millenium Institute of Astrophysics (MAS), by Fondecyt Regular 1150345 and by the BASAL-CATA Center for
Astrophysics and Associated Technologies PFB-06.
      
\end{acknowledgements}

%
%
\bibliographystyle{aa} 
\bibliography{biblio.bib}

\begin{thebibliography}{27}
\expandafter\ifx\csname natexlab\endcsname\relax\def\natexlab#1{#1}\fi

\bibitem[{{Bensby} {et~al.}(2013){Bensby}, {Yee}, {Feltzing}, {Johnson},
  {Gould}, {Cohen}, {Asplund}, {Mel{\'e}ndez}, {Lucatello}, {Han}, {Thompson},
  {Gal-Yam}, {Udalski}, {Bennett}, {Bond}, {Kohei}, {Sumi}, {Suzuki}, {Suzuki},
  {Takino}, {Tristram}, {Yamai}, \& {Yonehara}}]{Bensby13}
{Bensby}, T., {Yee}, J.~C., {Feltzing}, S., {et~al.} 2013, \aap, 549, A147

\bibitem[{{Carretta} {et~al.}(2009){Carretta}, {Bragaglia}, {Gratton}, \&
  {Lucatello}}]{Eugenio09}
{Carretta}, E., {Bragaglia}, A., {Gratton}, R., \& {Lucatello}, S. 2009, \aap,
  505, 139

\bibitem[{{Di Matteo} {et~al.}(2015){Di Matteo}, {G{\'o}mez}, {Haywood},
  {Combes}, {Lehnert}, {Ness}, {Snaith}, {Katz}, \& {Semelin}}]{DiMatteo15}
{Di Matteo}, P., {G{\'o}mez}, A., {Haywood}, M., {et~al.} 2015, \aap, 577, A1

\bibitem[{{Di Matteo} {et~al.}(2014){Di Matteo}, {Haywood}, {G{\'o}mez}, {van
  Damme}, {Combes}, {Hall{\'e}}, {Semelin}, {Lehnert}, \& {Katz}}]{DiMatteo14}
{Di Matteo}, P., {Haywood}, M., {G{\'o}mez}, A., {et~al.} 2014, \aap, 567, A122

\bibitem[{{Eggen} {et~al.}(1962){Eggen}, {Lynden-Bell}, \& {Sandage}}]{ELS62}
{Eggen}, O.~J., {Lynden-Bell}, D., \& {Sandage}, A.~R. 1962, \apj, 136, 748

\bibitem[{{Fern{\'a}ndez-Trincado} {et~al.}(2016){Fern{\'a}ndez-Trincado},
  {Robin}, {Moreno}, {Schiavon}, {Garc{\'{\i}}a P{\'e}rez}, {Vieira}, {Cunha},
  {Zamora}, {Sneden}, {Souto}, {Carrera}, {Johnson}, {Shetrone}, {Zasowski},
  {Garc{\'{\i}}a-Hern{\'a}ndez}, {Majewski}, {Reyl{\'e}}, {Blanco-Cuaresma},
  {Martinez-Medina}, {P{\'e}rez-Villegas}, {Valenzuela}, {Pichardo}, {Meza},
  {M{\'e}sz{\'a}ros}, {Sobeck}, {Geisler}, {Anders}, {Schultheis}, {Tang},
  {Roman-Lopes}, {Mennickent}, {Pan}, {Nitschelm}, \& {Allard}}]{JoseModel}
{Fern{\'a}ndez-Trincado}, J.~G., {Robin}, A.~C., {Moreno}, E., {et~al.} 2016,
  \apj, 833, 132

\bibitem[{{Garc{\'{\i}}a P{\'e}rez} {et~al.}(2013){Garc{\'{\i}}a P{\'e}rez},
  {Cunha}, {Shetrone}, {Majewski}, {Johnson}, {Smith}, {Schiavon}, {Holtzman},
  {Nidever}, {Zasowski}, {Allende Prieto}, {Beers}, {Bizyaev}, {Ebelke},
  {Eisenstein}, {Frinchaboy}, {Girardi}, {Hearty}, {Malanushenko},
  {Malanushenko}, {Meszaros}, {O'Connell}, {Oravetz}, {Pan}, {Robin},
  {Schneider}, {Schultheis}, {Skrutskie}, {Simmonsand}, \&
  {Wilson}}]{AnaElia13}
{Garc{\'{\i}}a P{\'e}rez}, A.~E., {Cunha}, K., {Shetrone}, M., {et~al.} 2013,
  \apjl, 767, L9

\bibitem[{{Gilmore} {et~al.}(2012){Gilmore}, {Randich}, {Asplund}, {Binney},
  {Bonifacio}, {Drew}, {Feltzing}, {Ferguson}, {Jeffries}, {Micela}, \&
  et~al.}]{GES}
{Gilmore}, G., {Randich}, S., {Asplund}, M., {et~al.} 2012, The Messenger, 147,
  25

\bibitem[{{Gonzalez} {et~al.}(2015){Gonzalez}, {Zoccali}, {Vasquez}, {Hill},
  {Rejkuba}, {Valenti}, {Rojas-Arriagada}, {Renzini}, {Babusiaux}, {Minniti},
  \& {Brown}}]{Oscar15}
{Gonzalez}, O.~A., {Zoccali}, M., {Vasquez}, S., {et~al.} 2015, \aap, 584, A46

\bibitem[{{Gonz{\'a}lez Hern{\'a}ndez} \& {Bonifacio}(2009)}]{GHB09}
{Gonz{\'a}lez Hern{\'a}ndez}, J.~I. \& {Bonifacio}, P. 2009, \aap, 497, 497

\bibitem[{{Haywood} {et~al.}(2016){Haywood}, {Di Matteo}, {Snaith}, \&
  {Calamida}}]{Misha16}
{Haywood}, M., {Di Matteo}, P., {Snaith}, O., \& {Calamida}, A. 2016, \aap,
  593, A82

\bibitem[{{Hill} {et~al.}(2011){Hill}, {Lecureur}, {G{\'o}mez}, {Zoccali},
  {Schultheis}, {Babusiaux}, {Royer}, {Barbuy}, {Arenou}, {Minniti}, \&
  {Ortolani}}]{Van11}
{Hill}, V., {Lecureur}, A., {G{\'o}mez}, A., {et~al.} 2011, \aap, 534, A80

\bibitem[{{Martinez-Valpuesta} \& {Gerhard}(2013)}]{MartinezValpuesta13}
{Martinez-Valpuesta}, I. \& {Gerhard}, O. 2013, \apjl, 766, L3

\bibitem[{{McWilliam} \& {Rich}(1994)}]{McWilliamRich}
{McWilliam}, A. \& {Rich}, R.~M. 1994, \apjs, 91, 749

\bibitem[{{McWilliam} {et~al.}(2013){McWilliam}, {Wallerstein}, \&
  {Mottini}}]{McWilliam13}
{McWilliam}, A., {Wallerstein}, G., \& {Mottini}, M. 2013, \apj, 778, 149

\bibitem[{{Ness} {et~al.}(2014){Ness}, {Asplund}, \& {Casey}}]{Ness14}
{Ness}, M., {Asplund}, M., \& {Casey}, A.~R. 2014, \mnras, 445, 2994

\bibitem[{{Ness} {et~al.}(2013){Ness}, {Freeman}, {Athanassoula},
  {Wylie-de-Boer}, {Bland-Hawthorn}, {Asplund}, {Lewis}, {Yong}, {Lane}, \&
  {Kiss}}]{NessPop}
{Ness}, M., {Freeman}, K., {Athanassoula}, E., {et~al.} 2013, \mnras, 430, 836

\bibitem[{{Randich} {et~al.}(2013){Randich}, {Gilmore}, \& {Gaia-ESO
  Consortium}}]{GESRandich}
{Randich}, S., {Gilmore}, G., \& {Gaia-ESO Consortium}. 2013, The Messenger,
  154, 47

\bibitem[{{Rojas-Arriagada} {et~al.}(2016{\natexlab{a}}){Rojas-Arriagada},
  {Recio-Blanco}, {de Laverny}, {Mikolaitis}, {Matteucci}, {Spittoni}, \&
  {Schultheis}}]{Alvaro2}
{Rojas-Arriagada}, A., {Recio-Blanco}, A., {de Laverny}, P., {et~al.}
  2016{\natexlab{a}}, \aap, submitted

\bibitem[{{Rojas-Arriagada} {et~al.}(2016{\natexlab{b}}){Rojas-Arriagada},
  {Recio-Blanco}, {de Laverny}, {Schultheis}, {Mikolaitis}, {Kordopatis},
  {Hill}, {Gilmore}, {Randich}, {Alfaro}, {Bensby}, {Koposov}, {Costado},
  {Franciosini}, {Hourihane}, {Jofr{\'e}}, {Lardo}, {Lewis}, {Lind}, {Magrini},
  {Monaco}, {Morbidelli}, {Sacco}, {Worley}, {Zaggia}, \&
  {Chiappini}}]{AlvaroDisc}
{Rojas-Arriagada}, A., {Recio-Blanco}, A., {de Laverny}, P., {et~al.}
  2016{\natexlab{b}}, \aap, 586, A39

\bibitem[{{Rojas-Arriagada} {et~al.}(2014){Rojas-Arriagada}, {Recio-Blanco},
  {Hill}, {de Laverny}, {Schultheis}, {Babusiaux}, {Zoccali}, {Minniti},
  {Gonzalez}, {Feltzing}, {Gilmore}, {Randich}, {Vallenari}, {Alfaro},
  {Bensby}, {Bragaglia}, {Flaccomio}, {Lanzafame}, {Pancino}, {Smiljanic},
  {Bergemann}, {Costado}, {Damiani}, {Hourihane}, {Jofr{\'e}}, {Lardo},
  {Magrini}, {Maiorca}, {Morbidelli}, {Sbordone}, {Worley}, {Zaggia}, \&
  {Wyse}}]{Alvaro1}
{Rojas-Arriagada}, A., {Recio-Blanco}, A., {Hill}, V., {et~al.} 2014, \aap,
  569, A103

\bibitem[{{Scannapieco} \& {Tissera}(2003)}]{ScannapiecoTissera03}
{Scannapieco}, C. \& {Tissera}, P.~B. 2003, \mnras, 338, 880

\bibitem[{{Schiavon} {et~al.}(2017{\natexlab{a}}){Schiavon}, {Johnson},
  {Frinchaboy}, {Zasowski}, {M{\'e}sz{\'a}ros}, {Garc{\'{\i}}a-Hern{\'a}ndez},
  {Cohen}, {Tang}, {Villanova}, {Geisler}, {Beers}, {Fern{\'a}ndez-Trincado},
  {Garc{\'{\i}}a P{\'e}rez}, {Lucatello}, {Majewski}, {Martell}, {O'Connell},
  {Prieto}, {Bizyaev}, {Carrera}, {Lane}, {Malanushenko}, {Malanushenko},
  {Mu{\~n}oz}, {Nitschelm}, {Oravetz}, {Pan}, {Roman-Lopes}, {Schultheis}, \&
  {Simmons}}]{Schiavon16b}
{Schiavon}, R.~P., {Johnson}, J.~A., {Frinchaboy}, P.~M., {et~al.}
  2017{\natexlab{a}}, \mnras, 466, 1010

\bibitem[{{Schiavon} {et~al.}(2017{\natexlab{b}}){Schiavon}, {Zamora},
  {Carrera}, {Lucatello}, {Robin}, {Ness}, {Martell}, {Smith},
  {Garc{\'{\i}}a-Hern{\'a}ndez}, {Manchado}, {Sch{\"o}nrich}, {Bastian},
  {Chiappini}, {Shetrone}, {Mackereth}, {Williams}, {M{\'e}sz{\'a}ros},
  {Allende Prieto}, {Anders}, {Bizyaev}, {Beers}, {Chojnowski}, {Cunha},
  {Epstein}, {Frinchaboy}, {Garc{\'{\i}}a P{\'e}rez}, {Hearty}, {Holtzman},
  {Johnson}, {Kinemuchi}, {Majewski}, {Muna}, {Nidever}, {Nguyen}, {O'Connell},
  {Oravetz}, {Pan}, {Pinsonneault}, {Schneider}, {Schultheis}, {Simmons},
  {Skrutskie}, {Sobeck}, {Wilson}, \& {Zasowski}}]{Schiavon16}
{Schiavon}, R.~P., {Zamora}, O., {Carrera}, R., {et~al.} 2017{\natexlab{b}},
  \mnras, 465, 501

\bibitem[{{Schultheis} {et~al.}(2016){Schultheis}, {Rojas-Arriagada}, {Garc\'ia
  P\'erez}, {Jonsson}, {Hayden}, {Nandakumar}, {Cunha}, \&
  {APOGEE}}]{Mathias16}
{Schultheis}, M., {Rojas-Arriagada}, A., {Garc\'ia P\'erez}, A.~E., {et~al.}
  2016, \aap, submitted

\bibitem[{{Sumi} {et~al.}(2004){Sumi}, {Wu}, {Udalski}, {Szyma{\'n}ski},
  {Kubiak}, {Pietrzy{\'n}ski}, {Soszy{\'n}ski}, {Wo{\'z}niak},
  {{\.Z}ebru{\'n}}, {Szewczyk}, \& {Wyrzykowski}}]{Sumi}
{Sumi}, T., {Wu}, X., {Udalski}, A., {et~al.} 2004, \mnras, 348, 1439

\bibitem[Zoccali et al.(2016)]{2016arXiv161009174Z} Zoccali, M., Vasquez, S., Gonzalez, O.~A., et al.\ 2016, arXiv:1610.09174

\end{thebibliography}

\begin{appendix}

\section{List of stars with low-$\alpha$ abundances}
The list of stars identified in Sect. 3 as having anomalously low-$\alpha$ abundances is included in Table A.1.

\begin{table*}
\caption{GES identificator, \teff, \logg, \feh, \mgfeabun, \sifeabun \ and $\Delta/\sigma$\aabun \ for the identified stars with \aabun \ underabundance. The reported uncertainties correspond to the measurement 
dispersion between the different analysis methods.}
\centering
{\scriptsize
\begin{tabular}{c c c c c c c c c}
\hline\hline 
cname    & \teff & \logg & \feh & \mgfeabun & \sifeabun & \alfeabun & $\Delta/\sigma$\aabun & setup configuration \\  
\hline
  18032412-3003001 & 4674 $\pm$  36 &   2.39 $\pm$   0.14 &   -0.66  $\pm$   0.03 &    0.24 $\pm$   0.02 &    0.26 $\pm$   0.02 &    --- &   2.0 & HR10,HR21     \\
  18034317-3006349 & 4819 $\pm$  95 &   2.65 $\pm$   0.35 &    -0.38 $\pm$   0.02 &    0.12 $\pm$   0.06 &    0.17 $\pm$   0.05 &    0.23 $\pm$   0.06  &   2.5 & HR10,HR21     \\
  17534571-4105165 & 4354 $\pm$  37 &   2.06 $\pm$   0.24 &    -0.39 $\pm$   0.02 &    0.07 $\pm$   0.06 &    0.18 $\pm$   0.02 &     ---                &   2.9 &  HR21     \\
  17553025-4106299 & 4606 $\pm$ 168 &   2.70 $\pm$   0.26 &    -0.58 $\pm$   0.02 &   -0.16 $\pm$   0.08 &    0.21 $\pm$   0.08 &    ---               &  5.3 &  HR21     \\
  17562638-4129329 & 4561 $\pm$  53 &   2.45 $\pm$   0.42 &    -0.33 $\pm$   0.05 &    0.12 $\pm$   0.08 &    0.15 $\pm$   0.07 &   --- &   2.4 & HR21     \\
  17571064-4145311 & 4959 $\pm$  87 &   2.74 $\pm$   0.50 &    -0.38 $\pm$   0.08 &    0.14 $\pm$   0.02 &    0.19 $\pm$   0.07 &    0.22 $\pm$   0.05 &   2.2 & HR21     \\
  17571166-4128521 & 4553 $\pm$  49 &   2.55 $\pm$   0.28 &    -0.06 $\pm$   0.05 &    0.02 $\pm$   0.09 &   -0.04 $\pm$   0.06 &    0.12 $\pm$   0.02 &   2.8 & HR21     \\
  17573538-4136125 & 4809 $\pm$  72 &   2.48 $\pm$   0.24 &    -0.80 $\pm$   0.02 &    0.23 $\pm$   0.02 &    0.11 $\pm$   0.09 &    0.16 $\pm$   0.03 &   3.6 & HR21     \\
  17582234-3449464 & 4560 $\pm$ 150 &   2.51 $\pm$   0.59 &     0.07 $\pm$   0.04 &   -0.06 $\pm$   0.08 &   -0.08 $\pm$   0.02 &    --- &  2.6 &  HR21    \\ 
  18030869-2956389 & 4607 $\pm$ 105 &   2.78 $\pm$   0.45 &    -0.11 $\pm$   0.07 &    0.03 $\pm$   0.09 &    0.04 $\pm$   0.03 &    0.26 $\pm$   0.04 &   2.4 & HR21     \\
  18033088-3003563 & 4551 $\pm$  43 &   2.71 $\pm$   0.08 &     0.41 $\pm$   0.02 &   -0.16 $\pm$   0.05 &   -0.21 $\pm$   0.02 &    0.00 $\pm$   0.03 &   2.1 & HR21    \\ 
  18035188-3005558 & 4574 $\pm$  65 &   2.86 $\pm$   0.28 &    -0.21 $\pm$   0.08 &    0.11 $\pm$   0.02 &    0.00 $\pm$   0.09 &    0.21 $\pm$   0.03 &   3.0  & HR21    \\ 
  18233838-3413015 & 4625 $\pm$  64 &   2.98 $\pm$   0.29 &     0.38 $\pm$   0.07 &   -0.22 $\pm$   0.03 &   -0.17 $\pm$   0.02 &   -0.07 $\pm$   0.04 &   2.4  & HR21   \\  
  18244002-2449329 & 5000 $\pm$ 216 &   2.83 $\pm$   0.52 &    -1.31 $\pm$   0.08 &    0.29 $\pm$   0.03 &    0.31 $\pm$   0.02 &   --- &   2.8 & HR21   \\  
  18250175-2514127 & 4463 $\pm$  91 &   2.79 $\pm$   0.08 &     0.51 $\pm$   0.02 &   -0.18 $\pm$   0.08 &   -0.29 $\pm$   0.04 &    0.02 $\pm$   0.02 &   2.3 & HR21    \\ 
  18251816-2529497 & 4663 $\pm$  45 &   2.78 $\pm$   0.37 &    -0.36 $\pm$   0.07 &  0.16 $\pm$   0.04 &     0.20 $\pm$   0.03 &    --- &   2.0 &HR21   \\ 
  18352178-2659169 & 4636 $\pm$  57 &   2.85 $\pm$   0.33 &   0.06 $\pm$   0.03 &   0.00 $\pm$   0.02 &   -0.04 $\pm$   0.04 & 0.17 $\pm$   0.06 &   2.0 &HR21   \\ 
  18360308-2716364 & 4664 $\pm$  36 &   2.43 $\pm$   0.28 &    -0.14 $\pm$   0.06 &    0.09 $\pm$   0.03 &    0.04 $\pm$   0.08 &    --- &   2.2 & HR21      \\
\hline
\end{tabular}
}
\end{table*}

\section{Visual inspection of spectral lines}
Figure B.1 presents two examples of $\alpha$-element lines (the CaII line at 8498.0 \AA and the MgI line at 8806.7 \AA) allowing to compare
the spectra of two stellar twins. The abundances differences in Ca and Mg are reported in the figure.

\begin{figure}[ht]
\includegraphics[width=9cm,height=4cm]{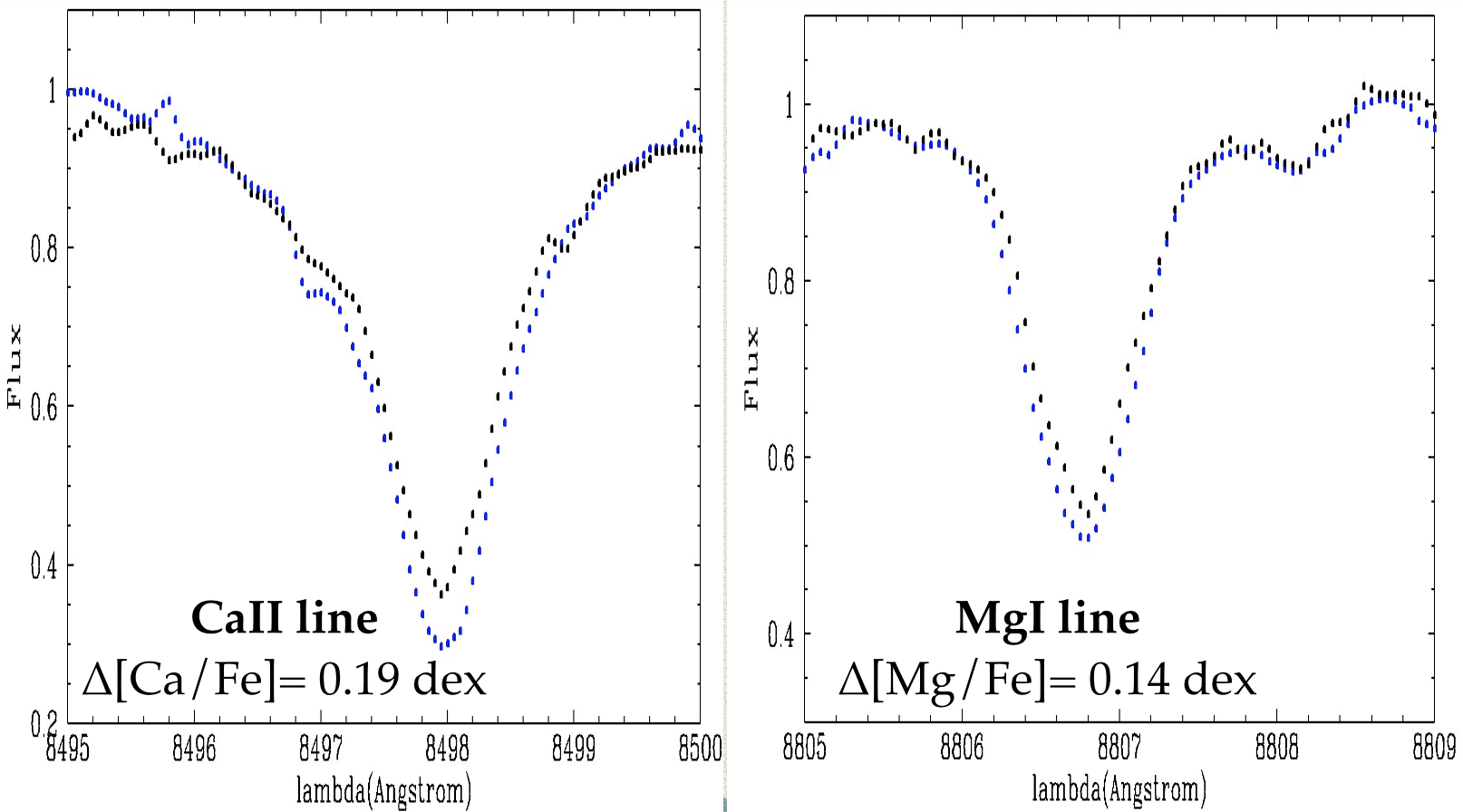} 
\caption{Comparison of the spectra of two stellar twins: the low-$\alpha$ star 18034317-3006349 (black points) and the star  18035553-3002562 with a standard \aabun \ composition (blue points)}
\label{Alrich}
\end{figure}

Figure B.2 shows two aluminum lines of one Al-rich star of the globular cluster NGC6522.

\begin{figure}[ht]
\includegraphics[width=9cm,height=4cm]{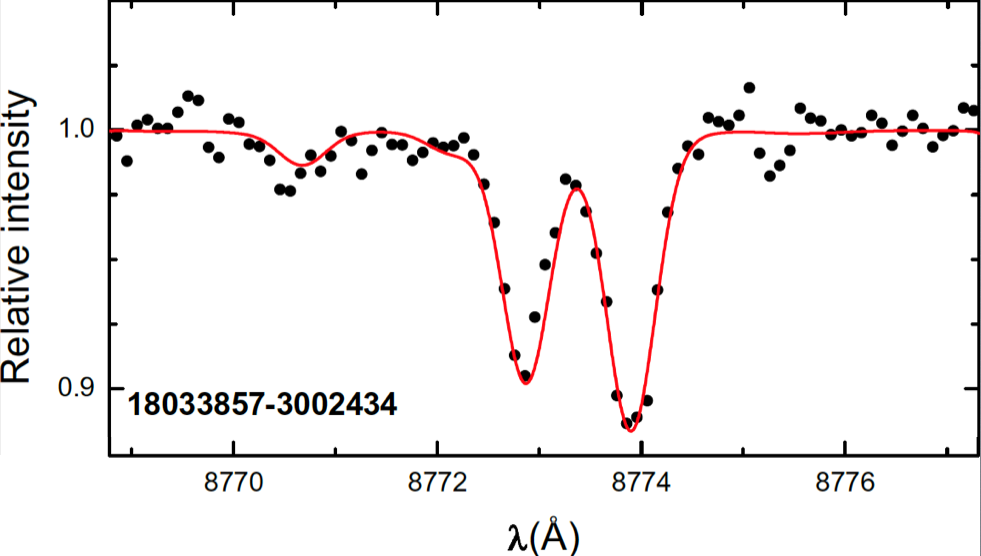} 
\caption{Spectrum of the star 18033857-3002434 (black points), 
around the two Al lines at 8772.8 and 8773.8 \AA. The red line shows a fit with a synthetic spectrum corresponding to \alfeabun=$0.78\pm$0.03~dex.}
\label{Alrich}
\end{figure}

\end{appendix}

\end{document}